\documentclass{jps-cp}
\usepackage{txfonts} %Please comment out this line unless the txfonts package is availabe in your LaTeX system.
\usepackage{braket}
\usepackage{wrapfig}
\newcommand{\Qhat}{\hat{Q}}
\newcommand{\Phat}{\hat{P}}
\newcommand{\Hhat}{\hat{H}}
\title{Nuclear structure and reaction with quantum shape fluctuation}
\author{Takashi \textsc{Nakatsukasa}$^{1,2,3}$, 
Yu \textsc{Kashiwaba}$^2$,
Fang \textsc{Ni}$^2$,
Kouhei \textsc{Washiyama}$^{1,4}$,
Kai \textsc{Wen}$^1$,
and
Nobuo \textsc{Hinohara}$^{1,2}$
}

\inst{$^{1}$Center for Computational Sciences, University of Tsukuba, Tsukuba 305-8577, Japan\\
$^{2}$Faculty of Pure and Applied Sciences, University of Tsukuba, Tsukuba 305-8577, Japan\\
$^{3}$RIKEN Nishina Center, Wako 351-0198, Japan\\
$^{4}$Department of Physics, Kyushu University, Fukuoka 819-0395, Japan\\
}

\recdate{August 2, 2019}

\abst{
We present recent results in theoretical studies on nuclear structure 
and reaction beyond mean field, using the adiabatic self-consistent
collective coordinate method and its extension.
We also present new results with the
finite-temperature Hartree-Fock-Bogoliubov calculation with
the three-dimensional-coordinate-space representation.
}

\begin{document}
\maketitle

\section{Introduction}

In studies of above-barrier nuclear fusion and giant resonances,
the time-dependent Hartree-Fock (TDHF) and time-dependent Hartree-Fock-Bogoliubov
(TDHFB) theories have been playing an important role \cite{Neg82,Sim12,NMMY16}.
One of great advantages of these approaches is to provide a unified non-empirical tool for microscopic
calculation of nuclear structure and reaction.
Recently, there are significant developments in the real-time simulation
\cite{Sim12,NMMY16,SY16,UOS16,BMRS16,MSW17,Sek17,USY17}.
The theory is capable of describing microscopic nuclear dynamics in the main channel.
However, since a part of quantum fluctuation is neglected
in these mean-field theories,
it is difficult to describe minor channels of the reaction.
Consequently, they fail to describe sub-barrier nuclear fusion, for instance.
As an example of nuclear structure problems,
we observe difficulty in description of transitional nuclei
in which the nuclear shape is fluctuating and not well defined.
An extension to ``beyond mean field'' is highly desired.

%The mean-field models for nuclear structure also have similar problems associated with
%missing quantum fluctuations.
%The problems are especially serious in transitional regions where the nuclear shape is fluctuating 
%and not well defined.
%The state should be expressed as a superposition of mean-field states with different shapes.
%The essential point is analogous to that in the sub-barrier fusion reaction:
%The theory lacks a part of important quantum fluctuation in the nuclear shape and
%in the relative motion between projectile and target.
%In order to recover the missing quantum fluctuation,
%theories beyond mean field are required.

A conventional way of improving the mean-field theory is
the generator coordinate method (GCM) \cite{RS80}.
The state is given by a superposition of (generalized) Slater determinants.
It is fully quantum mechanical and produces exact eigenstates in the limit of infinite number of
generator coordinates.
However, the GCM has some serious difficulties.
First of all, the generator-coordinate states are overcomplete in general.
The norm matrix in the generalized eigenvalue problem has eigenvalues
very close to zero, or even negative in practice.
In order to avoid numerical instability, we need to remove these
zero-norm eigenvectors which is computationally demanding and requires
a careful treatment.
Furthermore, there is no guarantee that dynamical effects are properly
taken into account by superposing only time-even Slater determinants.
In the case of center-of-mass motion in the GCM,
the use of complex generator coordinates is necessary to reproduce
the correct total mass,
suggesting that a pair of collective coordinates and momenta
should be simultaneously treated as generator coordinates
\cite{PT62,RS80}.
Another problem is associated with the use of effective interactions.
The effective interaction is designed to use in a given {\it truncated} space.
In this sense, the exact ground state in the {\it infinite} full space
is unphysical, or even not well defined.
Therefore,
we must carefully choose a small number of generator coordinates so as not
to produce the unphysical states.
This also prevents us from utilizing a variational method
to determine the generator coordinates \cite{SONY06,Fuk13}.
The GCM is also known to have a singular behavior when we use the
effective interaction which depends on a fractional power of one-body density
\cite{AER01,DSNR07,Dug09}.

In this paper, we use TDHF(B) trajectories on a collective subspace
to take into
account the quantum fluctuation beyond mean field.
A key ingredient of the method is a non-trivial task
to define the collective subspace which is decoupled from the other intrinsic
degrees of freedom, according to the TDHF(B) dynamics at low energy.
For this purpose, we utilize the 
adiabatic self-consistent collective coordinate (ASCC) method
\cite{MNM00,HNMM07,HNMM09,Nak12,NMMY16},
that is recapitulated in section \ref{sec:ASCC}.
The ASCC method consists of a set of equations to define the
collective subspace.
The solution of those equations provides a series of time-even
Slater determinants $\ket{\phi(q,p=0)}$,
and the generators of collective coordinates
and momenta $(\hat{Q}(q),\hat{P}(q))$ to span the collective submanifold;
$\ket{\phi(q,p)}\approx e^{ip\hat{Q}(q)}\ket{\phi(q,p=0)}$.
Using these, we calculate physical observables.
Two different methods are presented in section \ref{sec:requantization}.
Finally, in section \ref{sec:3D-HFB},
we present our recent developments of a new computer code
for the finite-temperature Hartree-Fock-Bogoliubov (HFB)
calculation in the three-dimensional coordinate space, which enables us
to study various problems in nuclear dynamics of finite nuclei and neutron stars.

%%%%%%%%%
\section{Adiabatic self-consistent collective coordinate (ASCC) method}
\label{sec:ASCC}

We assume that the TDHF(B) state $\ket{\phi(q,p)}$ is represented by 
a few collective canonical variables $(q,p)=(q^1,\cdot,q^K;p_1,\cdots,p_K)$.
Here, we assume that the number of collective variables are $2\times K$.
In the adiabatic limit of $p=0$, we denote $\ket{\phi(q,p=0)}$ as $\ket{\phi(q)}$.
The local generators of the collective variables $(p,q)$ are defined as
\begin{equation}
\Qhat^i(q)\ket{\phi(q)}=-i\frac{\partial}{\partial p_i}\ket{\phi(q)},
\quad\quad
\Phat_i(q)\ket{\phi(q)}= i\frac{\partial}{\partial q^i}\ket{\phi(q)}.
\end{equation}
Using the generator, we may express the TDHF(B) state $\ket{\phi(q,p)}$ as
%\begin{equation}
$
 \ket{\phi(q,p)}  =  \exp\left\{ i p_i \Qhat^i(q) \right\}
 \ket{\phi(q)} ,
$
%\end{equation}
based on the adiabatic time-even Slater determinants $\ket{\phi(q)}$.
Assuming that the momenta $p_i$ are rather small, we expand the basic
equations in terms of $p_i$ up to the second order, which leads to
the ASCC equations.

The ASCC method consists of the following three equations,
as the zeroth, first, and second order in momenta,
respectively \cite{MNM00,NMMY16}.
\begin{eqnarray}
&&\delta\bra{\phi(q)}\Hhat_M(q)\ket{\phi(q)} = 0,
\label{eq:mfHFB}
\\
&&\delta\bra{\phi(q)} \left[\Hhat_M(q), \Qhat^i(q)\right]
- \frac{1}{i} B^{ij}(q) \Phat_j(q)
+\frac{1}{2}\left[\frac{\partial V}{\partial q^j}\Qhat^j(q),
 \Qhat^i(q)\right]
\ket{\phi(q)} = 0,
\label{eq:ASCC1}
\\
&&\delta\bra{\phi(q)} \left[\Hhat_M(q),
 \frac{1}{i}\Phat_i(q)\right] - C_{ij}(q) \Qhat^j(q)
- \frac{1}{2}\left[\left[\Hhat_M(q), \frac{\partial V}{\partial
 q^k}\Qhat^k(q)\right], B_{ij}(q) \Qhat^j(q)\right]
\ket{\phi(q)} = 0, 
\label{eq:ASCC2}
\end{eqnarray}
where the indices $i$, $j$, and $k$ run through the number of collective
variables ($1,\cdots,K$).
$\Hhat_M(q)$ represents the moving Hamiltonian
\begin{equation}
  \Hhat_M(q) = \Hhat
 -  \frac{\partial V}{\partial q^i}\Qhat^i(q) ,
\end{equation}
and
%\begin{eqnarray}
\begin{equation}
 C_{ij}(q) = \frac{\partial^2 V}{\partial q^i \partial q^j} - \Gamma_{ij}^k\frac{\partial V}{\partial q^k},
\quad\quad
 \Gamma_{ij}^k(q) = \frac{1}{2} B^{kl}\left( \frac{\partial B_{li}}{\partial q^j}
 + \frac{\partial B_{lj}}{\partial q^i} - \frac{\partial B_{ij}}{\partial q^l} \right).
%\end{eqnarray}
\end{equation}
These equations are the basic equations for the ASCC method.
The solutions of the ASCC equations (\ref{eq:mfHFB}), (\ref{eq:ASCC1}), and
(\ref{eq:ASCC2}) provide the states $\ket{\phi(q)}$,
the generators $\Qhat(q)$ and $\Phat(q)$,
the collective potentials $V(q)$,
and the inertial masses $B_{ij}(q)$.
$B^{ij}(q)$ is a reciprocal inertial tensor, $B^{ik}B_{kj}=\delta^i_j$.
The HF(B) ground state corresponds to one of the states $\ket{\phi(q)}$ with $q=q_0$.
For details, the readers are referred to Ref.~\cite{NMMY16}
and references therein.
In the next section, we present requantization methods
to calculate physical observables and wave functions.

%%%%%%%%%%%%%%%%
\section{Requantization methods}
\label{sec:requantization}

In order to obtain information on stationary states (energy eigenstates),
the requantization of TDHF(B) is necessary.
We provide here two different approaches to this problem.

%%%%%%%%%%%%%%%%%%%
\subsection{Collective Hamiltonian approach}

Since the ASCC provides us the potential $V(q)$ and the inertial mass $B_{ij}(q)$
as functions of the collective coordinate $q$,
we may construct the {\it classical} collective Hamiltonian as
\begin{equation}
\mathcal{H}(q,p)=\frac{1}{2}B^{ij}(q) p_i p_j + V(q) .
	\label{eq:classical_H}
\end{equation}
Then, a possible way of construction of the quantum Hamiltonian is based on the Pauli's prescription.
\begin{equation}
	\hat{H}(q,p)=-\frac{1}{2}\frac{1}{\sqrt{\mathcal{B}(q)}} \frac{\partial}{\partial q^i}
	\sqrt{\mathcal{B}(q)} B^{ij}(q) \frac{\partial}{\partial q^j}
	+ V(q) ,
\end{equation}
where $\mathcal{B}\equiv \det\{B^{ij}\}$.

\begin{wrapfigure}{r}{0.45\textwidth}
\vspace{-20pt}
%	\centerline{
\includegraphics[width=0.45\textwidth]{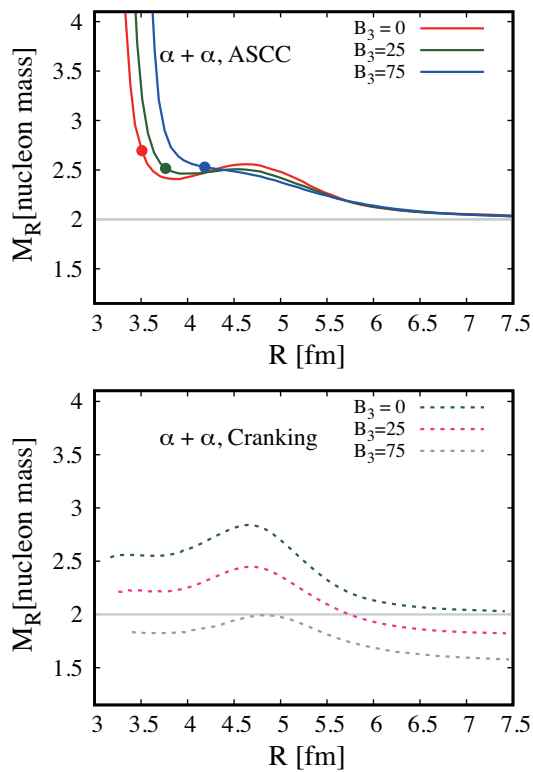}
%\includegraphics[width=0.45\textwidth]{M-R.pdf}
%}
\caption{
Inertial mass calculated with the ASCC (top panel)
and with the cranking formula (bottom).
The values of $B_3$ change from 0 to 75 MeV fm$^5$.
See text for details.
}
\label{fig:M-R}
	\vspace{-20pt}
\end{wrapfigure}
Now, let us show some recent results along this strategy.
The full ASCC solutions require large number of iterations to determine the collective subspace.
To facilitate this calculation, we neglect the last terms of equations
(\ref{eq:ASCC1}) and (\ref{eq:ASCC2}).
This leads to a set of equations similar to the quasiparticle-random-phase approximation (QRPA).
A self-consistent solution of equations (\ref{eq:mfHFB}), (\ref{eq:ASCC1}), and (\ref{eq:ASCC2}),
defines a state $\ket{\phi(q)}$ on the collective subspace with a certain value of $q$,
in addition to the local generators $(\Qhat^i(q), \Phat_i(q))$, the potential $V(q)$,
and the inertial masses $B_{ij}(q)$.
In order to define these quantities on the whole collective subspace spanned
by different values of $q$,
we need to repeat the calculation many times.
We apply this method to fusion reaction of light nuclei, using the BKN interaction
\cite{BKN76}.
In this case, we assume a single collective coordinate ($K=1$).
The main results with the BKN functional have been already published
in references \cite{WenN16,WenN17,WenN18}, in which
we have calculated the optimal reaction paths, scattering phase shifts,
and the astrophysical $S$-factors.
In this paper, in order to demonstrate an advantage of the method, we show our recent results
on the inertial mass for an {\it extended} BKN interaction.

The BKN interaction has no derivative terms.
Therefore, the nucleons' effective mass is identical to the bare nucleon mass ($m^*=m$).
However, most of realistic effective interactions have effective mass different from
the bare mass, typically $m^*/m \approx 0.7$.
In such cases, an improper treatment of collective motion leads to a wrong answer.
For instance, the translational mass of a nucleus is deviated from the total mass $Am$,
in the cranking formula and the Gaussian overlap approximation \cite{RS80}.
The time-odd mean fields are known to play a key role to restore the Galilean symmetry.

In order to check the capability of the ASCC method, we show in Fig.~\ref{fig:M-R}
the calculated collective mass for $\alpha+\alpha$ scattering.
Here, we consider adding a derivative term to the effective interaction.
Since the BKN interaction usually assumes the isospin degeneracy, we add
the following $B_3$ term to the original BKN energy density functional,
$B_3 \{ \tau(\mathbf{r}) \rho(\mathbf{r}) - \mathbf{j}^2(\mathbf{r}) \}$,
where $\tau(\mathbf{r})$ is the isoscalar kinetic density and
$\mathbf{j}(\mathbf{r})$ is the isoscalar current density.
The part of $\tau(\mathbf{r}) \rho(\mathbf{r})$ induces the effective mass $m^*$, while
the time-odd-density part of $\mathbf{j}^2(\mathbf{r})$ is necessary to recover the
Galilean symmetry.
With $B_3=0$, both the ASCC and the cranking formula reproduce the $\alpha+\alpha$ reduced mass
in the asymptotic limit $(R\rightarrow\infty$).
However, the cranking mass formula fails to do so with $B_3\neq 0$.
In contrast, the ASCC inertial mass converges to the correct reduced mass
no matter what $B_3$ value is.
This means that the ASCC equations take into account the time-odd effect to fully recover 
the Galilean symmetry.

We expect similar effect of the time-odd mean fields
on the rotational moments of inertia and vibrational inertial tensors.
In order to perform the calculation with a realistic effective interaction,
we further replace $\Qhat^i(q)$ in equation (\ref{eq:mfHFB}) by 
the mass quadrupole operators.
This is equivalent to the constrained-Hartree-Fock-Bogoliubov-plus-local-QRPA
formalism \cite{HSNMM10}.
In Ref.~\cite{WasN18},
we have shown the rotational moments of inertia for $^{106}$Pd in the quadrupole deformation plane
$(\beta,\gamma)$,
calculated with the Skyrme interaction of SkM$^*$.
It clearly shows effect of time-odd mean fields,
increase of the moments of inertia from those of the cranking (Inglis-Belyaev) formula.
The enhancement amounts to $30-40$ \%, though this enhancement factor significantly
varies as a function of $(\beta,\gamma)$.

%%%%%%%%%%%%%%%%%%%
\subsection{Stationary phase approximation of the path integral}

Another method of requantization of the TDHF(B) dynamics is based on
the stationary phase approximation (SPA) for the path integral formulation of the quantum
mechanics.
The SPA corresponds to the classical trajectory in general.
In the present case, this corresponds to the TDHF(B) trajectory.
In order to obtain the stationary states (energy eigenstates), we need to find TDHF(B) periodic
trajectories, which is a very difficult task in practice.
However, using the ASCC method, we may reduce the TDHF(B) phase space of huge dimension into
a collective subspace of small dimension.
In case of $K=1$, the dynamics in the subspace become integrable.
For the integrable systems, a semiclassical quantization method based on the SPA
is available \cite{KS80,Kur81,SM88}.
In order to take into account the quantum fluctuation beyond mean field,
recently we have adopted this new approach of ASCC+SPA.

We use the ASCC method to identify the collective subspace spanned by the collective variables
$(q,p)$.
Each point of the collective subspace corresponds to a generalized Slater determinant
$\ket{\phi(q,p)}$.
Then, we solve the classical Hamilton's equation of motion for the Hamiltonian (\ref{eq:classical_H}).
It is easy to find a periodic trajectory
because the system is practically one-dimensional ($K=1$), namely
integrable.
Then, the quantum eigenstates are given by
\begin{equation}
\ket{\psi_k}=\oint_{C_k} d\mu(q,p) \ket{\phi(q,p)} e^{i\mathcal{T}(q,p)} ,
	\label{eq:psi_k}
\end{equation}
where the closed trajectory $C_k$ satisfies the Einstein-Brillouin-Keller quantization rule,
\begin{equation}
	\oint_{C_k} pdq = 2\pi k,
	\quad\quad k=0, 1, 2, \cdots .
\end{equation}
Here, $d\mu(q,p)$ is an invariant measure, defined by
$
	\int d\mu(q,p) \ket{\phi(q,p)}\bra{\phi(q,p)} = 1
$, and
$\mathcal{T}(q,p)$ is the classical action,
\begin{equation}
	\mathcal{T}(q,p) = \int_0^t \braket{\phi(q(t'),p(t')) | i\frac{\partial}{\partial t'} 
	|\phi(q(t'),p(t'))} dt'  .
\end{equation}

The wave function of equation (\ref{eq:psi_k}) is nothing but superposition of many
Slater determinants on the collective subspace.
Advantages of this present method over the conventional GCM can be summarized as follows:
\newline
(i) The generator coordinate is not given by hand, but given by the TDHF(B) dynamics itself.
\newline
(ii) The generalized Slater determinants $\ket{\phi(q,p)}$ contain the time-odd
components (with $p\neq 0$).
\newline
(iii) No eigenvalue problem is involved.
\newline
\begin{wrapfigure}{r}{0.5\textwidth}
%	\vspace{-35pt}
%	\hspace{-35pt}
\includegraphics[width=0.5\textwidth]{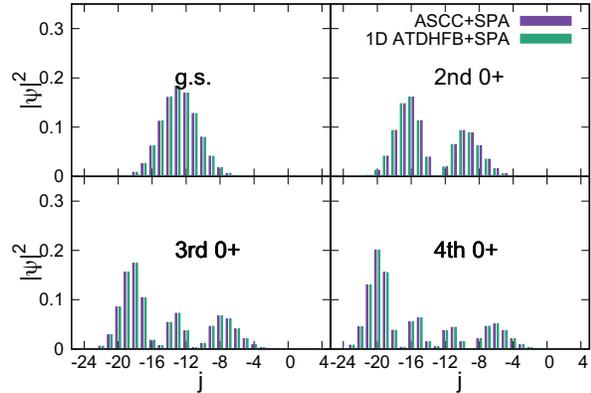}
%\includegraphics[width=0.35\textwidth,angle=-90]{tmp.eps}
%\includegraphics[width=0.35\textwidth, angle=-90]{N100Xeq2occ.pdf}
%	\vspace{30pt}
\caption{
Wave functions of the two-level pairing model with particle number of
$N=100$ and the pair strength $g=\Delta\epsilon/\Omega=0.2$,
where the level spacing is $\Delta\epsilon$ and
the degeneracy of each level is $2\Omega$ with $\Omega=50$.
See text and Ref.~\cite{NN18} for explanation.
	}
	\vspace{-15pt}
\label{fig:occ}
\end{wrapfigure}
Let us emphasize the last point (iii).
The GCM calculation requires us to solve the generalized eigenvalue problem (Hill-Wheeler equation).
This is a source of many difficulties in the GCM.
Instead, we solve the classical periodic trajectory for an integrable system, and
superpose the Slater determinants according to equation (\ref{eq:psi_k}).

We employ the pairing model to examine performance of the ASCC+SPA method
\cite{NN18,NHN18}.
We have shown the SPA method provides accurate results for
the multi-level pairing systems, which well agrees with the exact diagonalization of the Hamiltonian.
The particle-number projection is automatically performed by the superposition
(\ref{eq:psi_k}) along the classical trajectory of the pair rotation
\cite{NN18}.
In addition, with the ASCC, we can identify another classical trajectory
representing the pairing vibration.
The separation of these two collective modes is numerically performed in the ASCC.
Thus, here, we examine the ASCC's performance,
how well the pairing vibrational degrees of freedom are separated from
those of the pair rotation.
For the two-level case, this separation can be done analytically, and we compare the result
with the ASCC.
Figure~\ref{fig:occ} shows the obtained wave functions calculated with the ASCC+SPA and
the SPA with the TDHFB Hamiltonian expanded up to the second order in momenta
(denoted as ``1D ATDHFB+SPA'').
The wave function is decomposed into states labeled
by $j=n_2-n_1$ where $n_i$ represents the number of pairs in the $i$th level.
Thus, $j$ means half of the particle-number difference between the upper and the lower levels.
The agreement is almost perfect, which indicates that the separation between the pair rotation and
the pairing vibration is guaranteed in the ASCC.

%%%%%%%%%%%%%%%%%%%
\section{Hartree-Fock-Bogoliubov calculation on the three-dimensional coordinate space}
\label{sec:3D-HFB}

\begin{wrapfigure}{r}{0.3\textwidth}
	\vspace{-20pt}
\includegraphics[width=0.3\textwidth]{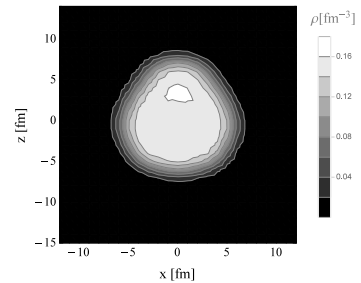}
	\vspace{-10pt}
\caption{
Nucleonic density distribution of $^{146}$Ba at $T=0$, calculated with SLy4 and
the volume pairing of the strength $g=-250$ MeV fm$^3$ and the cut-off $E_c=100$ MeV.
	}
\label{fig:146Ba}
\end{wrapfigure}
In order to achieve the full dynamics in shape/pairing fluctuation,
low-energy nuclear reaction, and crust structure of neutron stars,
a basic ingredient is the HFB calculation
that is able to compute nuclei with triaxial/octupole shapes,
and non-uniform crystalline structure.
In many cases, determining the stationary HFB solution
is even more difficult than the real-time TDHFB calculation.
Thus, developing an efficient computer program of the three-dimensional HFB calculation
is very important task for future.

Recently, an intriguing method of the HFB calculation has been proposed
\cite{JBRW17}.
The method does not require calculation of single-particle/quasi-particle
states.
Instead, the normal and pair densities $\rho$ and $\kappa$ are obtained
from the Green's function,
\begin{equation}
G(\mathbf{r}\sigma,\mathbf{r}'\sigma';E)
	=\begin{pmatrix}
		G_{UU}(\mathbf{r}\sigma,\mathbf{r}'\sigma';E) &
	G_{UV}(\mathbf{r}\sigma,\mathbf{r}'\sigma';E) \\
		G_{VU}(\mathbf{r}\sigma,\mathbf{r}'\sigma';E) &
	G_{VV}(\mathbf{r}\sigma,\mathbf{r}'\sigma';E)
	\end{pmatrix} ,
\end{equation}
with the integration over a contour in the complex-energy plane,
\begin{equation}
	\rho(\mathbf{r}\sigma,\mathbf{r}'\sigma') = \frac{1}{2\pi i}
	\oint_C  dz
	G_{VV}(\mathbf{r}\sigma,\mathbf{r}'\sigma';z) ,
	\quad\quad
	\kappa(\mathbf{r}\sigma,\mathbf{r}'\sigma') = \frac{1}{2\pi i}
	\oint_C  dz
	G_{UV}(\mathbf{r}\sigma,\mathbf{r}'\sigma';z) ,
\end{equation}
where the contour $C$ should contain the real axis of negative $z$,
up to the cut-off energy $z=-E_c$.
The Green's function is obtained by solving the equation,
$(z-H) G(\mathbf{r}\sigma,\mathbf{r}'\sigma';z)
= \delta(\mathbf{r}-\mathbf{r}')\delta_{\sigma\sigma'}$,
where $H$ is the HFB Hamiltonian.

We have extended this method for finite temperature and have developed a
computer code.
In Fig.~\ref{fig:146Ba},
we show an example of calculated density distribution for $^{146}$Ba at $T=0$ (HFB ground state).
It clearly shows a parity-violating octupole deformation, which is consistent with
former HFB+BCS calculation \cite{EN17}.
The proton pairing collapses, but the neutrons are in the superfluid phase with
the average gap of $\Delta_n=0.77$ MeV.
This is going to be an important step toward studies of nuclear
structure and reaction, including superfluid neutron dynamics in
the neutron-star crust \cite{KN19}.

%%%%%%%%%%%%%%%%%%%
%\section{Summary}
%\label{sec:summary}
\bigskip

\leftline{\bf Acknowledgement}

This work is supported in part by JSPS KAKENHI Grant No.18H01209
and No.19H05142,
and also by JSPS-NSFC Bilateral Program for Joint Research Project on
Nuclear mass and life for unravelling mysteries of r-process.
This research used computational resources of Oakforest-PACS,
through Multidisciplinary Cooperative Research Program
in Center for Computational Sciences, University of Tsukuba,
and the HPCI System Research Project (Project ID:hp190031).

\end{document}